\documentclass[pre,superscriptaddress,preprint]{revtex4}

\pdfoutput=1
\usepackage{psfrag}
\usepackage[T1]{fontenc}

\usepackage{amsfonts}
\usepackage{amssymb}

\usepackage{times}
\usepackage{pifont}
\usepackage{ae,aecompl}
\usepackage{amsmath}
\usepackage{graphicx}

\begin{document}

\title{Geometric Properties of the Three-Dimensional Ising and XY Models}
\author{Frank Winter} 
\altaffiliation{Present address: Deutsches Elektronen-Synchrotron DESY,
Platanenallee 6, 15738 Zeuthen, Germany}
\affiliation{Institut f\"ur Theoretische
Physik, Freie Universit\"at Berlin, Arnimallee 14, 14195 Berlin,
Germany} 
\author{Wolfhard Janke}
\affiliation{Institut f\"ur Theoretische Physik, Universit\"at Leipzig,
  Postfach 100920, 04009 Leipzig, Germany}
\author{Adriaan M. J. Schakel} \affiliation{Institut f\"ur Theoretische
Physik, Freie Universit\"at Berlin, Arnimallee 14, 14195 Berlin,
Germany}
\begin{abstract}
The fractal structure of high-temperature graphs of the
three-dimensional Ising and XY models is investigated by simulating
these graphs directly on a cubic lattice and analyzing them with the
help of percolation observables.  The Ising graphs are shown to percolate
right at the Curie critical point.  The diverging length scale relevant
to the graphs in the vicinity of the percolation threshold is shown to
be provided by the spin correlation length.  The fractal dimension of
the high-temperature graphs at criticality is estimated to be $D =
1.7349(65)$ for the Ising and $D = 1.7626(66)$ for the XY model.
\end{abstract}

\date{\today}

\maketitle

\section{Introduction}
The high-temperature (HT) expansion is a powerful tool to study the
critical properties of lattice spin models \cite{domb3}.  In this
approach, the partition function and correlation functions are
calculated by counting graphs on the lattice with each graph
representing a certain contribution.  Traditionally, such an expansion
is carried out exactly to a given order by enumerating all possible ways
a graph of given size and topology can be drawn on the lattice.  This
exact approach, involving combinatorial and graph-theoretical
algorithms, is notoriously challenging and laborious, with each
additional order requiring typically about the same amount of effort
needed for all previous orders combined.

\begin{figure}
\begin{center}
\input{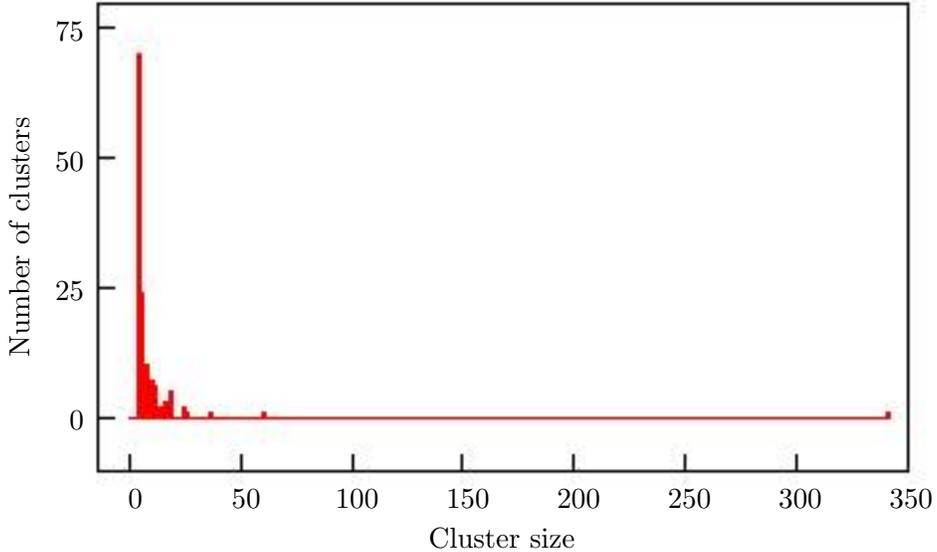} 
\end{center}
\caption{\label{fig:ns_I} Distribution of Ising HT graphs on
a cubic lattice of linear size $L=24$ at the percolation threshold.
Note the presence of a single big graph and many much smaller graphs.}
\end{figure}

We have developed a different approach \cite{geoPotts} in which the HT
representation of lattice spin models is studied by means of Monte Carlo
simulations.  The HT graphs along the links of the underlying lattice
are generated through a Metropolis plaquette update that proposes a
local change in the existing graph configuration.  At high temperatures,
only a few small graphs generated this way can be found scattered
throughout the lattice.  As the temperature is lowered, graphs start to
fill the lattice by growing larger and becoming more abundant.  At
temperatures below the critical temperature, the lattice becomes filled
with graphs.  A typical graph configuration now consists of one big
graph spanning the entire lattice and a collection of much smaller
graphs (see Fig.~\ref{fig:ns_I}).  The steady increase in the number of
occupied links and the appearance of graphs spanning the lattice as the
temperature is lowered are reminiscent of a percolation process.  The
use of percolation observables therefore suggests itself to analyze the
graph configurations.  For these observables to have bearing on the
critical properties of the model under investigation, it is necessary
that the HT graphs percolate right at the thermal critical point.  For
the Ising model on a square lattice we numerically showed that the
percolation threshold indeed coincides with the (exactly known) Curie
temperature \cite{geoPotts}.  In other words, the phase transition in
this lattice spin model to the ordered, low-temperature state manifests
itself through a proliferation of HT graphs.  Moreover, the fractal
structure of closed and open graphs was shown to encode the standard
critical exponents \cite{geoPotts,ht,ProkofevSvistunov,anomalous}.

The purpose of this paper is to extend this geometric Monte Carlo
approach to three dimensions.  Two-dimensional (2D) spin models arguably
form a very special class of models, in particular the Ising model as it
is self-dual.  It therefore is \textit{a priori} not obvious that this
geometric Monte Carlo approach together with the use of percolation
observables is viable in 3D.

Another Monte Carlo algorithm for studying HT representations of
classical statistical models has been put forward by Prokof'ev and
Svistunov \cite{worm}.  That so-called worm algorithm is much more
efficient than the conventional local update we use.  The dynamic
exponent $z$ characterizing the divergence of the autocorrelation time
$\tau$ when the critical point is approached, $\tau \sim \xi^z$, with
$\xi$ the correlation length, is close to zero for the worm algorithm
while it is larger than two for the plaquette update we use.  The
plaquette update, on the other hand, has the virtue that it provides a
direct and clean implementation of the HT representations of the spin
models we consider.  Prokof'ev and Svistunov \cite{ProkofevSvistunov}
recently applied their algorithm to the 3D complex $|\phi|^4$ theory to
determine the fractal dimension of the HT graphs in that theory.  These
graphs are unlike those in the XY model as links and vertices carry
different weights in the two models.  Moreover, because the update
algorithms differ, the graphs in the two models are simulated in
completely different ways.  Nevertheless, since the $|\phi|^4$ theory is
in the same universality class as the XY model, both types of HT graphs
should yield the same fractal dimension.  We set out below to
investigate whether universality holds for the fractal structure of HT
graphs.

The paper is organized as follows.  The next section introduces the
Metropolis plaquette update algorithm used in this Monte Carlo study
together with the percolation observables applied to analyze the HT
graphs.  The subsequent two sections present our results for the Ising
(Sec.~\ref{sec:Ising}) and XY (Sec.~\ref{sec:XY}) models, and the paper
ends in Sec.~\ref{sec:concl} with a summary and conclusions.

\section{Simulation and Data Analysis Techniques}
\label{sec:Simulation}
To be specific, we consider the HT representation of O($N$) lattice spin
models described by the Hamiltonian
\begin{equation} 
\label{H}
H = - J \sum_{\langle \mathbf{x}, \mathbf{x}' \rangle}
\mathbf{S}_\mathbf{x} \cdot \mathbf{S}_{\mathbf{x}'},
\end{equation} 
with the interaction, characterized by the parameter $J$, restricted to
spins on nearest-neighbor sites, so that the sum in Eq.~(\ref{H}) runs
only over nearest-neighbor pairs.  The spin variable
$\mathbf{S}_\mathbf{x}=(S^1_\mathbf{x}, \cdots , S^N_\mathbf{x})$
located at each site $\mathbf{x}$ of the cubic lattice has a fixed
length, $\mathbf{S}_\mathbf{x}^2 = 1$.   Simulations are carried out for
the 3D Ising ($N=1$) and XY ($N=2$) models.
\subsection{Ising Model}
The HT representation of the 3D Ising model on a cubic lattice 
with periodic boundary conditions consisting
of $N$ sites and $3N$ links \cite{Stanley},
\begin{equation} 
\label{HTIsing}
Z = (\cosh \beta)^{3N} 2^N
\sum_{\substack{\mathrm{closed} \\ \mathrm{graphs}}} K^b,
\end{equation}
provides an alternative, but completely equivalent description of the
spin model.  In this representation, which is purely geometric in
nature, spin degrees of freedom are swapped for link variables.  The
representation (\ref{HTIsing}) of the partition function can be
visualized as a sum over all possible closed graphs that can be drawn on
the lattice.  Each occupied link carries a factor $K=\tanh\beta$, with
$\beta$ the inverse temperature, where for convenience the coupling
constant $J$ is set to unity.  In the entire temperature range $0 \leq
\beta \leq \infty$, $0 \leq K \leq 1$.  The minimum number of occupied
links $b$ needed to form a closed graph is four on a cubic lattice.  The
internal energy
\begin{equation}
\label{inte} 
E = - \frac{\partial \ln Z}{\partial \beta} = - 3NK - \frac{1}{\sinh
\beta \, \cosh \beta} \, \langle b \rangle
\end{equation} 
is determined by the average number $\langle b \rangle$ of occupied
links.

The central idea of our geometric Monte Carlo approach \cite{geoPotts}
is to directly simulate the graphs contributing to the partition
function.  The HT representation (\ref{HTIsing}) suggests the following
local Metropolis update algorithm \cite{Erkinger}.

The probability distribution $P(G)$ for a given graph configuration $G$
reads in equilibrium
\begin{equation} 
  P(G) = \frac{1}{Z} (\cosh \beta)^{3N} 2^N  K^b.
\end{equation} 
Such a configuration can be reached from the configuration present
after, say, $t$ iterations in the  following way
\begin{equation}
  P_{t+1}(G) = P_t(G) + \sum_{G'} \left[ P_t(G') W(G' \to G) -  
    P_t(G) W(G \to G') \right],
\end{equation} 
where $W(G \to G')$ is the probability for the system to move from the
graph configuration $G$ with $b$ occupied links to the graph
configuration $G'$ with $b'$ occupied links.  In equilibrium, $P_{t+1}(G)
= P_t(G) = P(G)$, and the system satisfies detailed balance
\begin{equation} 
  P(G') W(G' \to G)  =  P(G) W(G \to G'),
\end{equation} 
or
\begin{equation} 
\frac{W(G \to G')}{W(G' \to G)} = \frac{P(G')}{P(G)} = \frac{K^{b'}}{K^b} .
\end{equation} 
As is custom with Metropolis algorithms, the acceptance rate
$p_\mathrm{HT}$ of a proposed update is maximized by giving the largest
of the two transition probabilities $W(G \to G')$ and $W(G' \to G)$
appearing in the ratio the largest possible value, which is one.  That
is, if the number of links $b'$ in the proposed configuration is larger
than the number of links $b$ in the existing configuration, so that
$K^{b'}/{K^b}<1$, $W(G' \to G)=1$ and $W(G \to G')=p_\mathrm{HT}$.  If,
on the other hand, $b'<b$, the proposed configuration carries a larger
weight than the existing one and will always be accepted.  

\begin{figure}
\begin{center}
\includegraphics[width=0.25\textwidth]{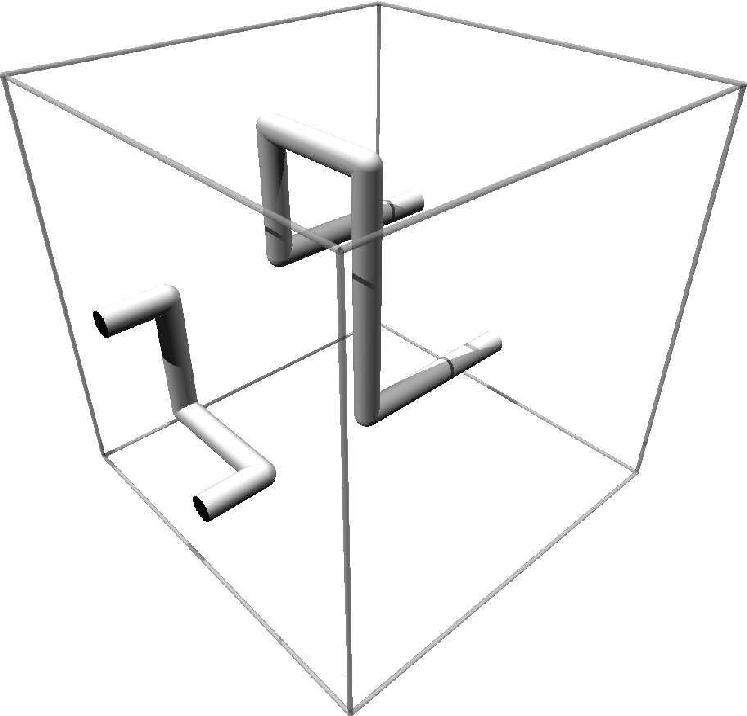} \hspace{.05\textwidth}
\includegraphics[width=0.25\textwidth]{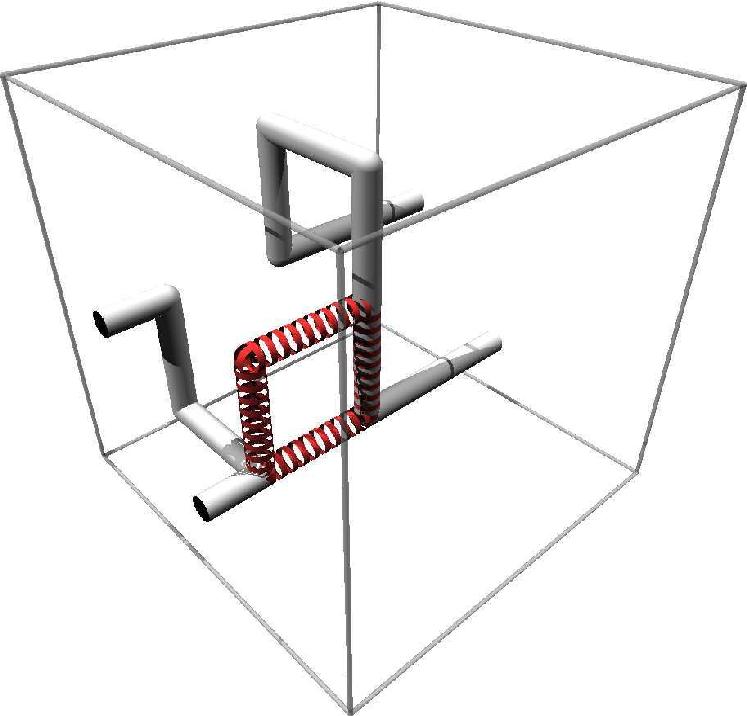} \hspace{.05\textwidth}
\includegraphics[width=0.25\textwidth]{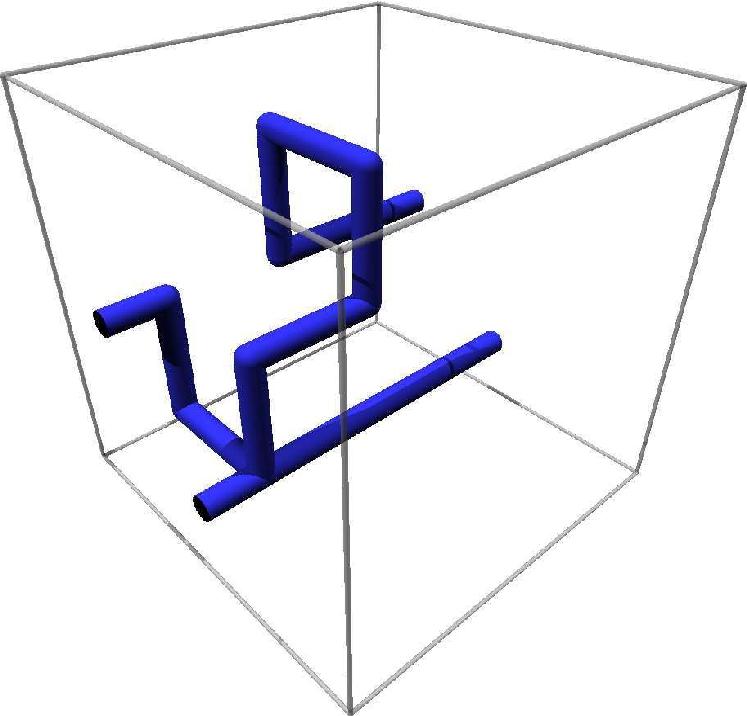} 
\end{center}
\caption{\label{fig:plaq} An existing Ising HT graph on a cubic lattice
(left panel) is updated with the help of a chosen plaquette (middle
panel) into a new graph (right panel).}
\end{figure}

The HT graphs are generated by taking the smallest possible closed
graphs on the lattice, i.e., plaquettes, as building blocks.  During a
sweep of the lattice, all plaquettes are visited in a regular,
typewriter fashion.  For the Ising model, reflecting the underlying
Z$_2$ spin symmetry, a link can either be empty or occupied.  The links
of a plaquette considered for update are changed from empty to occupied
and \textit{vice versa} (see Fig.~\ref{fig:plaq} for an illustration).
This is easily implemented by means of the binary rules $0+1=1,\;1+1=0$,
respectively.
The acceptance rate of a proposed update reads in formula \cite{Erkinger}
\begin{equation} 
\label{Metro}
p_\mathrm{HT} = \left\{ \begin{array}{ll} K^{b'-b} & \quad \mbox{if}
    \;\; \quad b'>b \\ 1 & \quad \mbox{else} . \end{array} \right. 
\end{equation} 
The resulting and existing number of occupied links, $b'$ and $b$,
respectively, are related via
\begin{equation} 
\label{ll}
b' = b + 4 - 2 b_\square,
\end{equation}  
with $b_\square$ denoting the number of links on the plaquette
already occupied.  By taking plaquettes as building blocks, the
resulting graphs are automatically closed.

Table~\ref{tabdsci1} gives a summary of the number $N_{\text{MC}}$ of
Monte Carlo sweeps of the lattice of size $L$ used for data collection
in the temperature interval $[\beta_1,\beta_2]$, with $N_{\text{cal}}$
sweeps used for equilibration.  The temperature intervals are sampled
with $i$ equidistant points.  The largest lattice taken for
temperature-dependent runs is $L=48$, while runs at the percolation
threshold are carried out on lattices up to $L=64$.  Since the plaquette
update is a local update, autocorrelation times grow very large on large
lattices.  Analyzing the time series of the percolation strength, we
estimate the autocorrelation time to vary from $\tau \approx 25$ for
$L=10$ to as long as $\tau \approx 2500$ for $L=64$.  To somewhat reduce
correlations between successive data points, we take measurements every
fifth sweep of the lattice.  Statistical errors are estimated by means
of jackknife binning.  Fits are carried out by using the nonlinear
Marquardt-Levenberg algorithm for minimization of error weighted least
squares.
%
%
\begin{table}
\caption{\label{tabdsci1} Overview of parameters used in the simulations
of the 3D Ising model.  The temperature intervals $[\beta_1,\beta_2]$
are sampled with $i$ equidistant points.  For each sampling point,
$N_{\text{MC}}$ Monte Carlo sweeps of the lattice of size $L$ are used
for data collection, with $N_{\text{cal}}$ sweeps used for
equilibration. }
\begin{ruledtabular}
\begin{tabular}{cccccc}
$\beta_1$ & $\beta_2$ & $i$ & $L$ &  $N_{\text{MC}}$ & $N_{\text{cal}}$ \\
\hline
$0.218000$ & $0.233000$ & $70$ & $12$ & $460000$ & $60000$ \\
$0.170000$ & $0.218000$ & $190$ & $12$ & $120000$ & $20000$ \\
$0.233000$ & $0.270000$ & $120$ & $12$ & $120000$ & $20000$ \\
\hline
$0.220000$ & $0.230000$ & $60$ & $16$ & $460000$ & $60000$ \\
$0.170000$ & $0.220000$ & $190$ & $16$ & $120000$ & $20000$ \\
$0.230000$ & $0.270000$ & $140$ & $16$ & $120000$ & $20000$ \\
\hline
$0.220500$ & $0.229000$ & $50$ & $20$ & $460000$ & $60000$ \\
$0.170000$ & $0.220500$ & $190$ & $20$ & $120000$ & $20000$ \\
$0.229000$ & $0.270000$ & $150$ & $20$ & $120000$ & $20000$ \\
\hline
$0.220500$ & $0.227100$ & $40$ & $24$ & $460000$ & $60000$ \\
$0.170000$ & $0.220500$ & $190$ & $24$ & $120000$ & $20000$ \\
$0.227100$ & $0.270000$ & $160$ & $24$ & $120000$ & $20000$ \\
\hline
$0.220500$ & $0.226000$ & $35$ & $28$ & $460000$ & $80000$ \\
$0.170000$ & $0.220500$ & $180$ & $28$ & $120000$ & $30000$ \\
$0.226000$ & $0.270000$ & $165$ & $28$ & $120000$ & $30000$ \\
\hline
$0.220500$ & $0.224900$ & $30$ & $32$ & $460000$ & $80000$ \\
$0.170000$ & $0.220500$ & $190$ & $32$ & $120000$ & $30000$ \\
$0.224900$ & $0.270000$ & $170$ & $32$ & $120000$ & $30000$ \\
\hline
$0.220500$ & $0.223800$ & $25$ & $40$ & $500000$ & $120000$ \\
$0.170000$ & $0.220500$ & $180$ & $40$ & $130000$ & $40000$ \\
$0.223800$ & $0.270000$ & $175$ & $40$ & $130000$ & $40000$ \\
\hline
$0.220500$ & $0.223500$ & $25$ & $48$ & $560000$ & $240000$ \\
$0.170000$ & $0.220500$ & $150$ & $48$ & $140000$ & $50000$ \\
$0.223500$ & $0.270000$ & $160$ & $48$ & $140000$ & $50000$ \\
\end{tabular}
\end{ruledtabular}
\end{table}

\subsection{XY Model}
For the XY model ($N=2$), where the spins take values along a circle,
the general spin Hamiltonian (\ref{H}) with $J=1$ reduces to
\begin{equation} 
H = - \sum_{\langle \mathbf{x}, \mathbf{x}' \rangle}
\cos(\theta_\mathbf{x} - \theta_{\mathbf{x}'} ),
\end{equation} 
with $\theta_\mathbf{x}$ being the planar angle of the spin at site
$\mathbf{x}$ relative to a fixed, but arbitrary axis.  The HT expansion
is facilitated by the use of the Fourier expansion
\begin{equation} 
\mathrm{e}^{\beta \cos(\theta)} = \sum_{l=-\infty}^\infty I_l(\beta)
\, \mathrm{e}^{\mathrm{i} l \theta},
\end{equation} 
where $I_l(x)$ is the modified Bessel function of the first kind.  The
partition function,
\begin{equation} 
Z = \prod_\mathbf{x} \left[\int_{-\pi}^\pi \frac{\mathrm{d}
\theta_\mathbf{x}}{2 \pi} \right] \prod_{\langle \mathbf{x}, \mathbf{x}'
\rangle} \mathrm{e}^{- \beta \cos(\theta_\mathbf{x} -
\theta_{\mathbf{x}'} )}\end{equation} 
then takes the well-known form \cite{Polyakov}
\begin{equation} 
Z = \prod_\mathbf{x} \left[\int_{-\pi}^\pi \frac{\mathrm{d}
\theta_\mathbf{x}}{2 \pi} \right] \prod_{\langle \mathbf{x}, \mathbf{x}'
\rangle} \, \sum_{l_{\mathbf{x}, \mathbf{x}'}} I_{l_{\mathbf{x},
\mathbf{x}'}}(\beta) \, \mathrm{e}^{\mathrm{i} l_{\mathbf{x},
\mathbf{x}'}(\theta_\mathbf{x} - \theta_{\mathbf{x}'} )},
\end{equation} 
involving the integers $l_{\mathbf{x}, \mathbf{x}'}$ defined on the
links connecting the nearest neighbor sites $\mathbf{x}$ and
$\mathbf{x}'$.  The spin degrees of freedom are now easily integrated
out with the result
\begin{equation} 
\label{zxy}
Z = \prod_{\langle \mathbf{x}, \mathbf{x}' \rangle} \,
\sideset{}{'}\sum_{l_{\mathbf{x}, \mathbf{x}'}} I_{l_{\mathbf{x},
\mathbf{x}'}}(\beta) = I_0^{3N}(\beta) \prod_{\langle \mathbf{x},
\mathbf{x}' \rangle} \, \left(1 + \sideset{}{'}\sum_{l_{\mathbf{x},
\mathbf{x}'} \neq 0} \frac{I_{l_{\mathbf{x},
\mathbf{x}'}}(\beta)}{I_0(\beta)} \right).
\end{equation} 
Here, the prime on the sums is to indicate that only configurations
satisfying the zero divergence condition, $\sum_{\mathbf{x}'}
l_{\mathbf{x}, \mathbf{x}'} = 0$ at each site $\mathbf{x}$ contribute,
where the sum $\sum_{\mathbf{x}'}$ runs over all nearest neighbors of
$\mathbf{x}$.  It is generally accepted that the link variables
$l_{\mathbf{x}, \mathbf{x}'}$ at the right hand of Eq.~(\ref{zxy}) can
be restricted to the values $\pm 1$ without changing the universality
class.  The partition function of the XY model can then be cast in a
form analogous to the HT representation (\ref{HTIsing}) of the Ising
model,
\begin{equation} 
Z = I_0^{3N}(\beta) \sum_{\substack{\mathrm{closed} \\ \mathrm{oriented}
\; \mathrm{graphs}}} K^b,
\end{equation} 
where $K \equiv I_1(\beta)/I_0(\beta)$, with $0 \leq K <1$ for all
$\beta$, and use is made of the property that $I_{-1}(x)=I_1(x)$.  Since
the link variable of the truncated model can take two nontrivial values
$\pm1$, the graphs now have, in contrast to the Ising model, an
orientation.  A plaquette considered for update must therefore also be
given a (randomly chosen) orientation.  As binary rules we now have
$-1+1=0,\; 0+1=1,\; 0-1=-1,\; 1-1=0$ in addition to the restrictions
$-1-1=0,\; 1+1=0$ of singly occupancy which, as already mentioned, we
expect not to change the universality class of the model.  Apart from
these modifications, the HT representation of the XY model can be
handled in the same way as that of the Ising model.
\subsection{Observables}
The HT graphs are analyzed with the help of standard percolation
observables \cite{StaufferAharony}.  An important characteristic is
whether a graph spans the lattice or not.  We say a graph does so
already if it spans the lattice in just one of the three possible
directions. By recording this each time the graphs are analyzed, one
obtains the percolation probability $P_\mathrm{S}$, which tends to zero
in the limit $\beta \to 0$ and to unity in the opposite limit $\beta \to
\infty$.  Another important observable is the graph distribution $n_b$,
which gives the average number of graphs of $b$ occupied links
normalized by the volume.  Close to the percolation threshold it assumes
the form
\begin{equation}
\label{nb} 
n_b \sim b^{-\tau} \mathrm{e}^{- \theta b}, \quad \quad \theta \propto
|K/K_\mathrm{per}-1|^{1/\sigma} .
\end{equation} 
The exponents $\tau$ and $\sigma$ are related to the fractal dimension
$D$ of the HT graphs via $\tau = d/D + 1$ with $d=3$ the dimension of
the space box, and \cite{ht}
\begin{equation}
\label{sigma} 
\sigma=1/\nu D, 
\end{equation} 
where $\nu$ is the exponent characterizing the divergence of the
correlation length $\xi$ as the percolation threshold is approached,
$\xi \propto |K/K_\mathrm{per}-1|^{-\nu}$.  An additional observable we
measure is the percolation strength $P_\infty$, which is defined as the
size of the largest graph normalized by the volume.
Finally, we also record the average graph size $\chi_\mathrm{G}$.  

The percolation threshold and the fractal dimension of the HT graphs
follow from applying finite-size scaling to these observables.
According to scaling theory, the percolation probability and strength
for different values of the tuning parameter $K$ and for different
lattice sizes $L$ do not depend on these variables separately, but
depend on them in a convoluted way \cite{BinderHeermann}
\begin{equation}  
\label{fss}
P_\mathrm{S}(K) = {\sf P}_\mathrm{S}\left[L^{1/\nu}
(K/K_\mathrm{per}-1)\right], \quad P_\infty (K) = L^{-\beta_\mathrm{G}/\nu}
\, {\sf P}_\infty\left[L^{1/\nu} (K/K_\mathrm{per}-1)\right] .
\end{equation} 
Here, $\beta_\mathrm{G}$ determines the scaling dimension of the
percolation strength, which is related to the fractal dimension $D$ of
the HT graphs through \cite{StaufferAharony}
\begin{equation} 
D = d - \beta_\mathrm{G}/\nu .
\end{equation} 
The scaling dimension of the percolation probability is zero.  The first
scaling relation in Eq.~(\ref{fss}) implies that the curves
$P_\mathrm{S}(K)$ measured on lattices of different size all cross at
the same point.  This point, being volume independent, marks the
percolation threshold $K_\mathrm{per}$ on the infinite lattice.  That
scaling relation in addition implies that the curves collapse onto a
universal curve when replotted as a function of $L^{1/\nu}
(K/K_\mathrm{per}-1)$.  Similarly, the second scaling relation in
Eq.~(\ref{fss}) implies that if in addition to the horizontal axis also
the vertical axis is properly rescaled, with the correct value for the
ratio $\beta_\mathrm{G}/\nu$, also the $P_\infty (K)$ data fall onto a
universal curve.  That scaling relation in addition implies that
measurements at the percolation threshold scale as $P_\infty
(K_\mathrm{per}) \propto L^{-\beta_\mathrm{G}/\nu}$, providing a good
means of determining $\beta_\mathrm{G}/\nu$.

\section{Ising Model Results}
\label{sec:Ising}
\begin{figure}
\begin{center}
\input{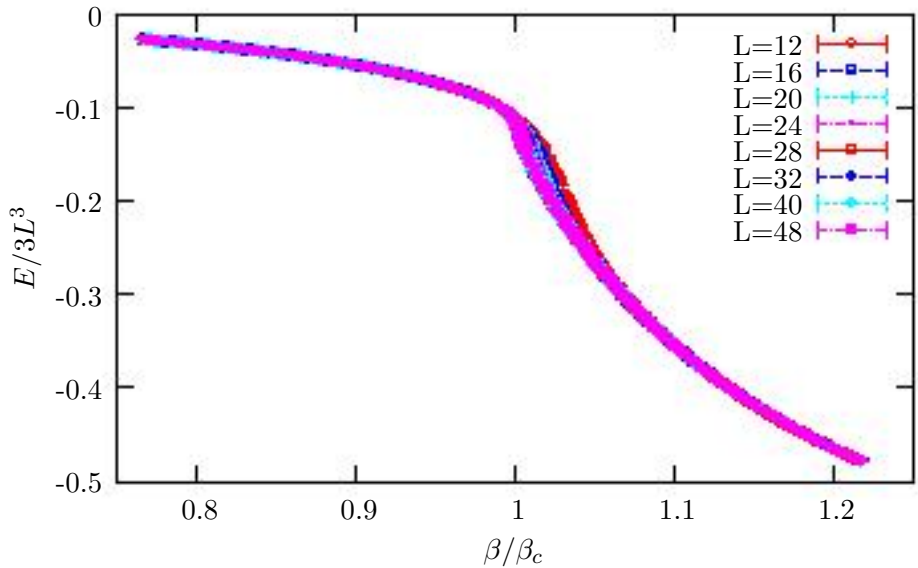}
\end{center}
\caption{\label{fig:e_I} Average Ising HT graph density as a function of
the reduced inverse temperature $\beta/\beta_\mathrm{c}$ for cubic
lattices varying in linear size from $L=12$ to $L=48$.}
\end{figure}

We start by simulating the HT graphs of the Ising model on a cubic
lattice.  Figure~\ref{fig:e_I} shows the internal energy as obtained
through Eq.~(\ref{inte}) by measuring the density of occupied links.
Apart from a small interval around the critical temperature, $E$ is seen
to be almost independent of volume, implying that the correlation length
is much smaller here than the linear size of the smallest lattice
considered ($L=12$).  When entering the critical region, the correlation
length becomes larger and eventually exceeds the size of the largest
lattice considered ($L=48$).

\begin{figure}
\begin{center}
\input{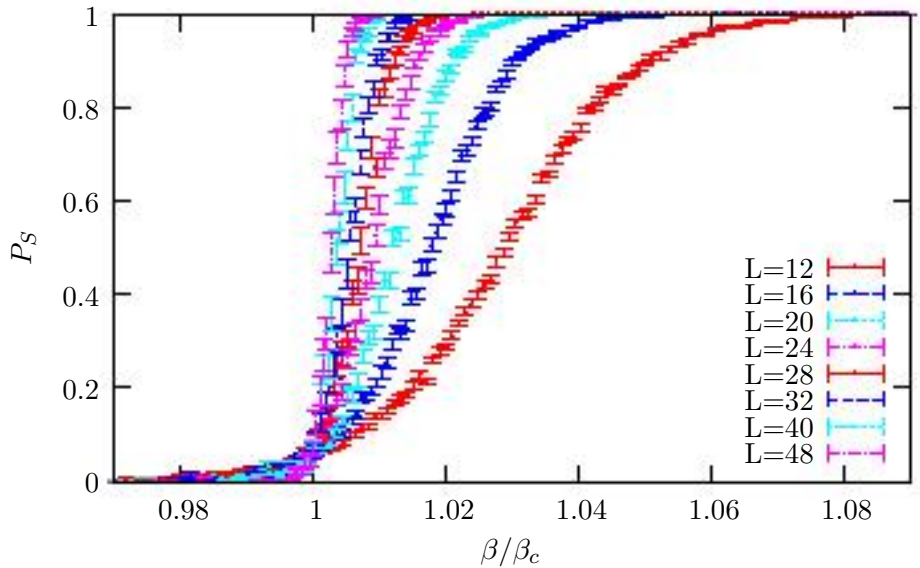}
\end{center}
\caption{\label{fig:pi_I} Percolation probability $P_\mathrm{S}$ of the
Ising HT graphs as a function of the reduced inverse temperature
$\beta/\beta_\mathrm{c}$ for cubic lattices varying in linear size from
$L=12$ to $L=48$.}
\end{figure}

\begin{figure}
\begin{center}
\input{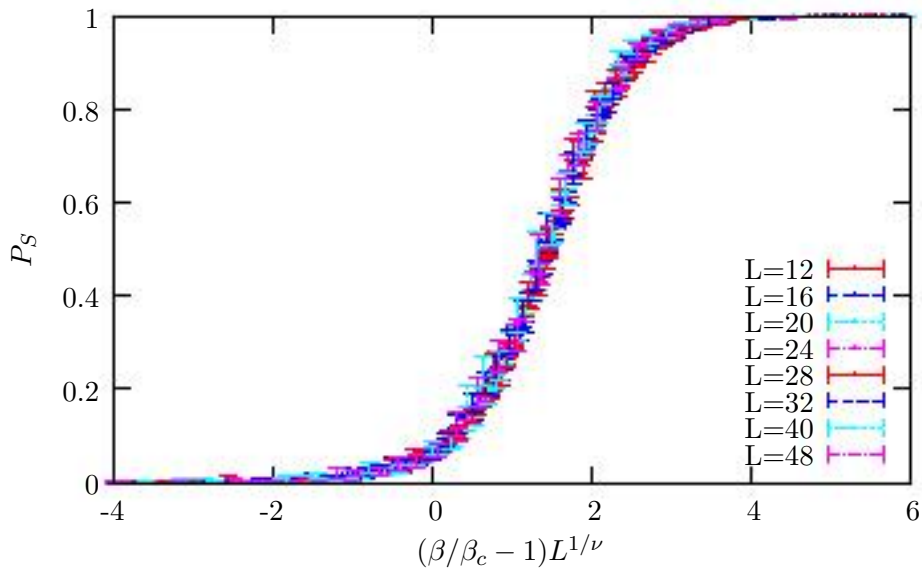}
\end{center}
\caption{\label{fig:pifss_I} Collapse of the data in Fig.~\ref{fig:pi_I}
with the choices (\ref{betanu}) for $\beta_\mathrm{c}$ and $\nu$.}
\end{figure}

We first determine the location of the percolation threshold. To
this end, we measure the percolation probability on lattices of different
linear size $L$ as a function of the inverse temperature (see
Fig.~\ref{fig:pi_I}).
The figure shows that the curves connecting the data points all cross
within error bars at the critical temperature, giving a first indication
that the percolation threshold coincides with the thermal critical
point.  The percolation probability at the threshold we estimate to be
$P_\mathrm{S}=0.05(2)$.  To obtain a more refined test, we apply
finite-size scaling to this observable.  When replotted as a function of
$(\beta/\beta_\mathrm{c}-1)L^{1/\nu}$, with $\nu$ the correlation length
exponent of the 3D Ising model, the data is expected to collapse onto a
universal curve.  The inverse critical temperature and $\nu$ have been
determined to high precision in Refs.~\cite{betacI,nuI}, 
\begin{equation} 
\label{betanu}
\beta_\mathrm{c} = 0.22165459(10) , \quad \nu = 0.63012(16) ,
\end{equation} 
respectively.  Figure~\ref{fig:pifss_I} shows that these values indeed
produce a good collapse of the data in the entire temperature range.
Since this is achieved without any adjustable parameter, we arrive at
the important conclusion that the percolation threshold of the HT graphs
coincides within error bars with the thermal critical point.  Moreover,
since the standard Ising correlation length exponent $\nu$ has been
used, it follows that the relevant diverging length scale for the HT
graphs in the vicinity of the critical temperature is provided by the
spin correlation length.

\begin{figure}
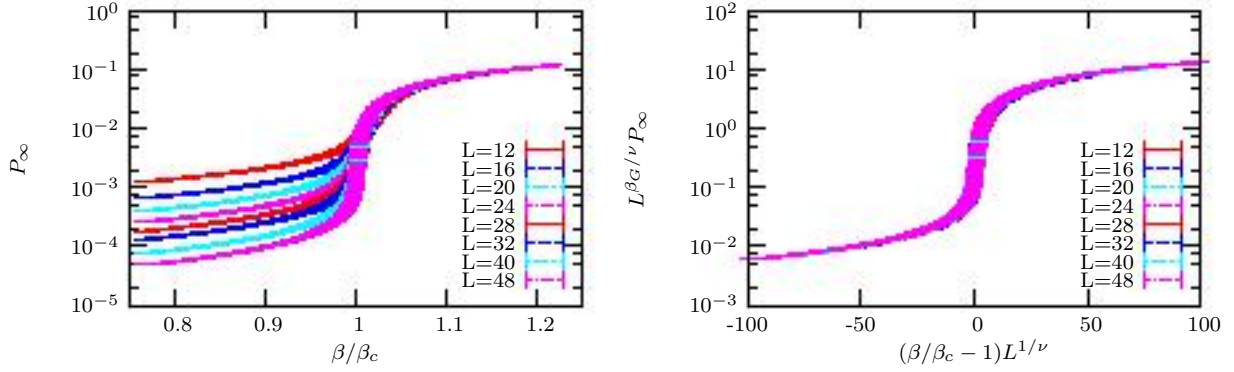

\begin{center}
\scriptsize
\input{p8_log_I.tex}
\input{p8fss1165_I.tex}
\end{center}
\caption{\label{fig:p8fss_I} Collapse of the Ising percolation strength data
measured on lattices of different size $L$ (left panel) when
$L^{\beta_\mathrm{G}/\nu} P_\infty$ is plotted as a function of
$(\beta/\beta_\mathrm{c}-1)L^{1/\nu}$ (right panel) on a semilogarithmic
scale.}
\end{figure}

We proceed to determine the fractal dimension $D$ of the HT graphs.
This can be done by measuring, for example, the percolation strength or
the average graph size at the percolation threshold and applying
finite-size scaling to the data obtained for different lattice sizes
$L$.  Since $\chi_\mathrm{G}$ was found to show large corrections to
scaling for small $L$, the observable $P_\infty$, showing only small
corrections, is used to estimate $D$.  Table~\ref{tabfitsi1lr}
summarizes the results of two-parameter fits using the nonlinear
Marquardt-Levenberg algorithm for various fit intervals.
%
%
\begin{table}
\caption{\label{tabfitsi1lr} Percolation strength exponent $\beta_G/\nu$
of the Ising HT graphs at the critical temperature (\ref{betanu}) as
obtained through two-parameter fits in the indicated intervals.}
\begin{ruledtabular}
\begin{tabular}{lll|lll}
 $L$ & $\beta_\mathrm{G}/\nu$ & $\chi^2/\mathrm{DOF}$ 
&$L$ & $\beta_\mathrm{G}/\nu$ & $\chi^2/\mathrm{DOF}$ \\ 
\hline 
 $8-64$ & $1.256 (5)$ & $1.61 $&$10-64$ & $1.265 (7)$ & $1.35 $ \\ 
 $8-56$ & $1.255 (6)$ & $1.68 $&$10-56$ & $1.264 (7)$ & $1.43 $ \\ 
 $8-48$ & $1.255 (6)$ & $1.75 $&$10-48$ & $1.266 (8)$ & $1.43 $ \\ 
 $8-40$ & $1.258 (6)$ & $1.57 $&$10-40$ & $1.271 (7)$ & $0.96 $ \\
 $8-32$ & $1.254 (7)$ & $1.57 $ &$10-32$ & $1.267 (8)$ & $1.05 $ \\ 
 $8-24$ & $1.253 (7)$ & $1.50 $ &$10-24$ & $1.269 (8)$ & $0.85 $ \\ 
 $8-20$ & $1.245 (8)$ & $1.19 $ &$10-20$ & $1.260 (11)$ & $0.83 $ \\
\hline
$12-64$& $1.256 (5)$ & $1.61$ & $14-64$ & $1.270 (10)$ & $1.41$  \\
$12-56$& $1.270 (9)$ & $1.41$ & $14-56$ & $1.269 (12)$ & $1.53$  \\
$12-48$& $1.273 (9)$ & $1.37$ & $14-48$ & $1.273 (13)$ & $1.51$  \\
$12-40$& $1.281 (7)$ & $0.68$ & $16-64$ & $1.261 (13)$ & $1.37$  \\
$12-32$& $1.279 (9)$ & $0.79$ & $16-56$ & $1.258 (15)$ & $1.47$  \\
$12-24$& $1.285 (7)$ & $0.34$ & $16-48$ & $1.261 (17)$ & $1.49$  \\
\end{tabular}                     
\end{ruledtabular}                
\end{table}                       
On the basis of these fits, we estimate the critical exponent
$\beta_\mathrm{G}/\nu$ to be
\begin{equation} 
\label{betaG}
\beta_\mathrm{G}/\nu = 1.2651 (65),
\end{equation} 
corresponding to the largest possible fit interval ($L=10-64$) that
still gives a good fit quality with a $\chi^2$ per degree of freedom
(DOF), $\chi^2/\mathrm{DOF} = 1.35$.
\begin{figure}
\begin{center}
\input{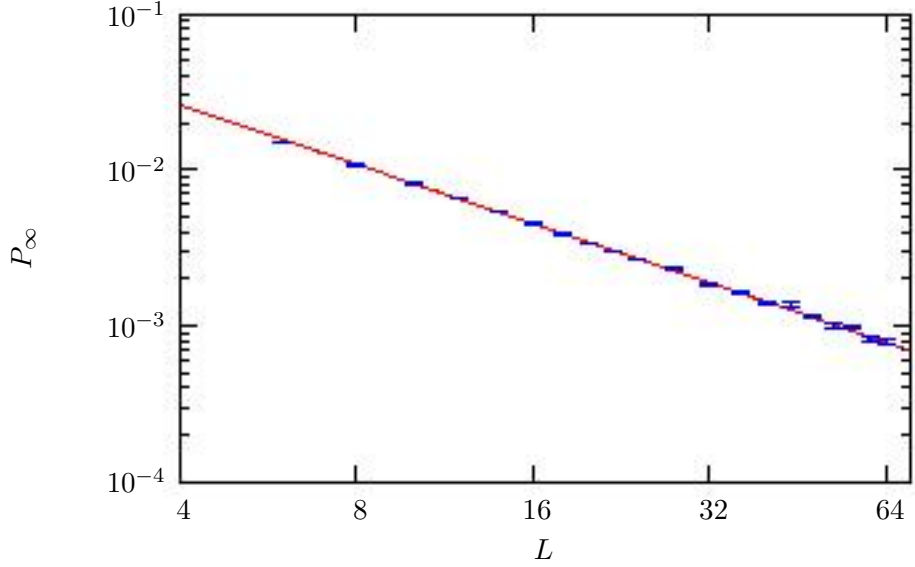}
\end{center}
\caption{\label{fig:p8_I} Log-log plot of the percolation strength
  $P_\infty$ of the Ising HT graphs at the critical temperature
  (\ref{betanu}) as a function of the lattice size $L$.  The straight
  line is obtained through a two-parameter fit in the interval
  $L=10-64$.}
\end{figure}
Figure~\ref{fig:p8fss_I} shows that, with this choice, the data collected
in the vicinity of the critical temperature on lattices of different
size $L$ fall onto a universal curve when both axes are properly
rescaled.

The estimate (\ref{betaG}) for the graph exponent leads to the estimate
\begin{equation} 
\label{DI}
D = d - \beta_\mathrm{G}/\nu = 1.7349(65)
\end{equation} 
($d=3$) for the fractal dimension of the HT graphs at the critical
point.  This fractal dimension is closer to that of a self-avoiding walk
in 3D, for which \cite{Guida} $D = 1/\nu = 1.7001(32)$, than to that of
a Brownian random walk, for which $D=2$.  Given this estimate for the
fractal dimension, the relation (\ref{sigma}) leads to the estimate
\begin{equation}
\label{sigmaI} 
\sigma = 0.9147(42)
\end{equation} 
for the graph distribution exponent $\sigma$, which for self-avoiding
random walks is unity.

\section{XY Model Results}
\label{sec:XY}
\begin{figure}
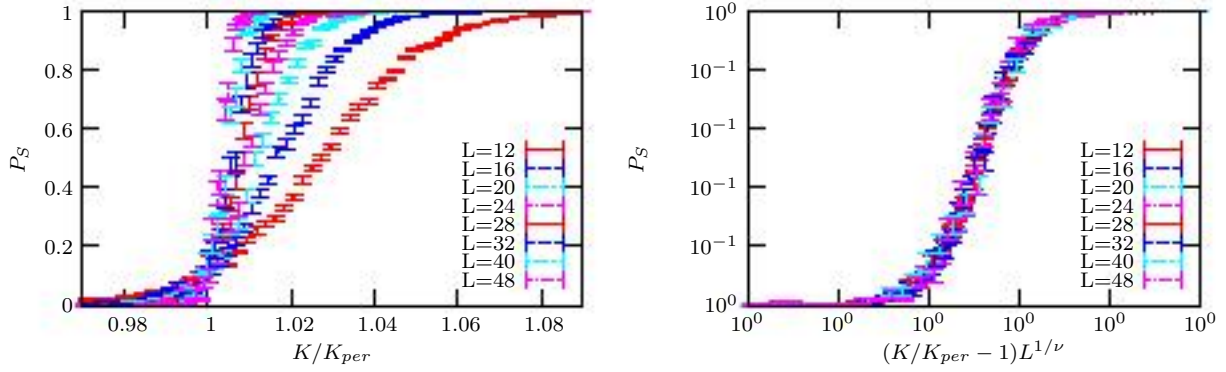

\begin{center}
\scriptsize
\input{pi_XY.tex}
\input{pifss_XY.tex}
\end{center}
\caption{\label{fig:pi_XY} Collapse of the percolation probability
$P_\mathrm{S}$ of XY HT graphs measured on lattices of different size
$L$ (left panel) when replotted as a function of
$(K/K_\mathrm{per}-1)L^{1/\nu}$ (right panel).}
\end{figure}

We proceed by analyzing the HT graphs of the 3D XY model.  The
critical temperature of the truncated XY model, where links are allowed
to be at most singly occupied, has to our knowledge not been determined
before.  To arrive at an accurate estimate of the percolation threshold,
we consider the percolation strength data and search for the best data
collapse given the value
\begin{equation} 
\nu = 0.6717(1)
\end{equation}
for the XY correlation length exponent recently reported in
Ref.~\cite{pisa}.  This is done by rendering a motion picture out of about
300 single frames showing the data collapse for different values of
$K_\mathrm{c}$. Successive frames correspond to slightly increased
values of $K_\mathrm{c}$.  A media player, such as MPlayer that can go
forward and backward frame by frame, is used to play the motion picture,
and to determine the value of $K_\mathrm{c}$ with the best collapse.
The quality of the collapse is established visually.  Error estimates are
based on the number of successive frames for which the quality of
collapse remains roughly the same.  We have checked this method by
applying it to the Ising model, where the critical temperature is known
to high precision, and obtained surprisingly good results.  For the
truncated XY model, we arrive in this way at the estimate
\begin{equation} 
\label{KperXY}
K_\mathrm{per} = 0.22288(5),
\end{equation} 
which is to be compared to the value $K_\mathrm{c} =
\tanh\beta_\mathrm{c} = 0.218095 \cdots $ of the Ising model.
Figure~\ref{fig:pi_XY} shows the collapse of the data achieved with the
estimate (\ref{KperXY}) of the percolation threshold.

\begin{figure}
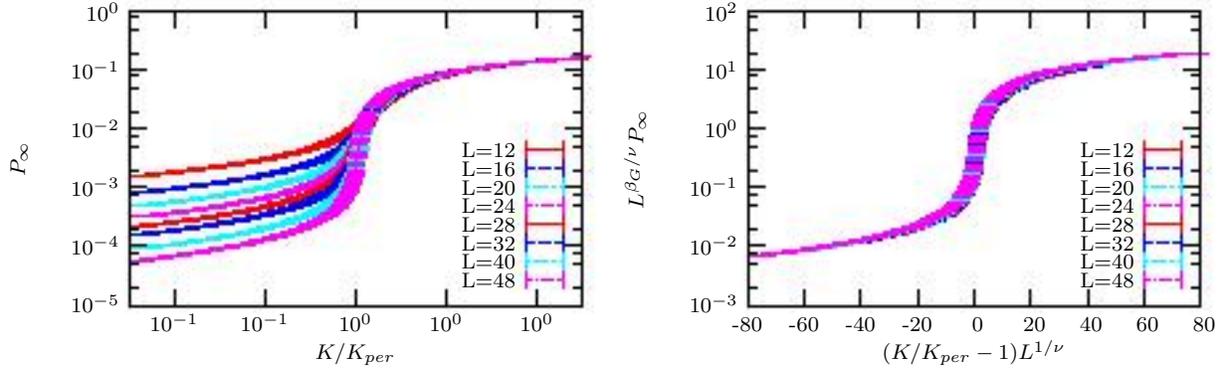

\begin{center}
\scriptsize
\input{p8_log_XY.tex}
\input{p8fss_lr_XY.tex}
\end{center}
\caption{\label{fig:p8fss_} Collapse of the XY percolation strength data
measured on lattices of different size $L$ (left panel) when
$L^{\beta_\mathrm{G}/\nu} P_\infty$ is plotted as a function of
$(K/K_\mathrm{per}-1)L^{1/\nu}$ (right panel) on a semilogarithmic
scale. }
\end{figure}

As for the Ising model, we determine the fractal dimension of the HT
graphs of the XY model by measuring the percolation strength $P_\infty$
at the percolation threshold (\ref{KperXY}) on lattices of different
size.  Table~\ref{tabfitsxylr} summarizes the results of two-parameter
fits using various fit intervals.
%
%
\begin{table} 
\caption{\label{tabfitsxylr} Percolation strength exponent $\beta_G/\nu$
of the XY HT graphs at the percolation threshold (\ref{KperXY}) as
obtained through two-parameter fits in the indicated
intervals.}
\begin{ruledtabular}
\begin{tabular}{lll|lll}
 $L$ & $\beta_G/\nu$   & $\chi^2/\mathrm{DOF}$ 
&$L$ & $\beta_G/\nu$   & $\chi^2/\mathrm{DOF}$ \\ 
\hline
 $6-64$   & $1.203 (7)$  & $4.03 $ & $8-64$   & $1.221 (7)$ & $2.34 $ \\
 $6-56$   & $1.200 (7)$  & $3.93 $ & $8-56$   & $1.217 (7)$ & $2.36 $ \\
 $6-48$   & $1.196 (6)$  & $3.21 $ & $8-48$   & $1.212 (7)$ & $1.85 $ \\
 $6-40$   & $1.192 (6)$  & $2.73 $ & $8-40$   & $1.208 (7)$ & $1.60 $ \\
 $6-32$   & $1.188 (7)$  & $2.66 $ & $8-32$   & $1.203 (8)$ & $1.71 $ \\
 $6-24$   & $1.185 (7)$  & $2.83 $ & $8-24$   & $1.200 (9)$ & $1.98 $ \\
 $6-20$   & $1.179 (8)$  & $2.35 $ & $8-20$   & $1.192 (11)$& $1.95 $ \\
\hline
 $L$ & $\beta_G/\nu$   & $\chi^2/\mathrm{DOF}$ 
&$L$ & $\beta_G/\nu$   & $\chi^2/\mathrm{DOF}$ \\ 
\hline
$10-64$   & $1.237 (7)$ & $1.21 $  & $12-64$ & $1.244 (8)$ & $1.13$ \\
$10-56$   & $1.235 (7)$ & $1.26 $  & $12-56$ & $1.241 (9)$ & $1.22$ \\
$10-48$   & $1.229 (6)$ & $0.88 $  & $12-48$ & $1.234 (9)$ & $0.89$ \\
$10-40$   & $1.224 (7)$ & $0.77 $  & $14-64$ & $1.258 (7)$ & $0.56$ \\
$10-32$   & $1.221 (8)$ & $0.90 $  & $14-56$ & $1.257 (8)$ & $0.53$ \\
$10-24$   & $1.220 (11)$ & $1.12$  & $16-64$ & $1.258 (9)$ & $0.60$ \\
$10-20$   & $1.213 (15)$ & $1.36$  & $16-56$ & $1.256 (10)$ &$0.68$ \\
\end{tabular}
\end{ruledtabular}
\end{table}
On the basis of these fits, we estimate the critical exponent
$\beta_\mathrm{G}/\nu$ to be
\begin{equation} 
\label{betaXY}
\beta_\mathrm{G}/\nu = 1.2374 (66),
\end{equation} 
corresponding to the largest possible fit interval ($L=10-64$) that
still gives a good fit quality ($\chi^2/\mathrm{DOF} = 1.21$).
\begin{figure}
\begin{center}
\input{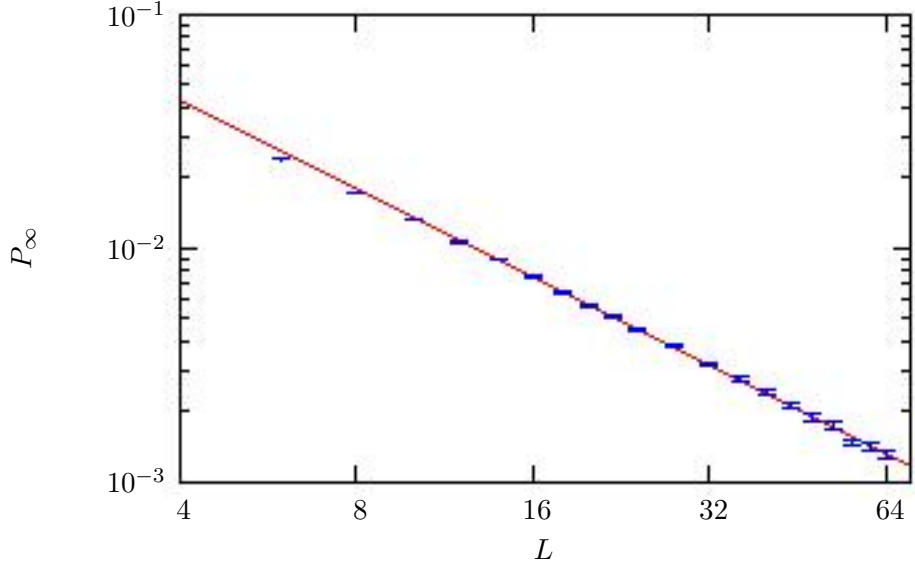}
\end{center}
\caption{\label{fig:p8_XY} Log-log plot of the percolation strength
  $P_\infty$ of the XY HT graphs at the percolation threshold
  (\ref{KperXY}) as a function of the lattice size $L$.  The straight
  line is obtained through a two-parameter fit in the interval
  $L=10-64$.}
\end{figure}
Figure~\ref{fig:p8fss_} shows that with this choice, the data collected
in the vicinity of the percolation threshold on lattices of different
size $L$ fall onto a universal curve when both axes are properly
rescaled.
The result (\ref{betaXY}) leads to the estimate
\begin{equation} 
\label{DXY}
D = 1.7626(66)
\end{equation}  
for the fractal dimension of XY HT graphs at the percolation threshold.
Comparison with the estimate (\ref{DI}) for the fractal dimension of the
Ising HT graphs shows that the XY graphs are more crumpled although
still much less so than a Brownian random walk.  For the graph
distribution exponent $\sigma$ we arrive at the estimate
\begin{equation} 
\sigma = 0.8446(45),
\end{equation} 
which is lower than the estimate (\ref{sigmaI}) for the Ising model.
Our estimate (\ref{DXY}) is in good agreement with the value $D =
1.7655(20)$ recently reported by Prokof'ev and
Svistunov~\cite{ProkofevSvistunov} for the $|\phi|^4$ theory which they
obtained using their worm algorithm~\cite{worm}.  As stated in the
Introduction, the HT graphs of that model as well as the worm update
algorithm used to simulate them are completely different from the XY HT
graphs and the plaquette update.  Yet, despite these differences, the
fractal dimensions of the two models, which share the same universality
class, agree within error bars.  We take this as a strong indication
that, as expected, universality holds for the fractal structure of HT
graphs.
\section{Conclusions}
\label{sec:concl}
In this paper, it was shown that the geometric Monte Carlo approach
originally introduced in the context of 2D spin models~\cite{geoPotts},
in which HT graphs are simulated directly and analyzed with the help of
percolation observables, can also be applied to 3D spin models.  The 3D
Ising HT graphs were shown to percolate right at the critical
temperature, which is known to high precision.  The phase transition to
the ordered, low-temperature state in this spin model was shown to be
reflected by a proliferation of HT graphs.  Also, through data collapse,
it was shown that the diverging length scale relevant to the HT graphs
in the vicinity of the percolation threshold is the spin correlation
length.  With the help of finite-size scaling, the fractal dimensions of
the closed Ising and XY HT graphs were determined as in percolation
theory.  Both models are handled similarly, with the only difference
that, in contrast to the Ising HT graphs, the XY HT graphs are oriented.
Finally, it was shown that universality holds for the fractal structure
of HT graphs.

\end{document}